\documentclass[12pt]{article}
\usepackage{pdproc,epsfig}
\usepackage{amssymb}

\def\ptps{{Prog.\ Theor.\ Phys.\ Suppl. \ }}
\newcommand{\be}{\begin{equation}}
\newcommand{\ee}{\end{equation}}
\newcommand{\bea}{\begin{eqnarray}}
\newcommand{\eea}{\end{eqnarray}}
\newcommand{\ba}{\begin{eqnarray}}
\newcommand{\ea}{\end{eqnarray}}
\newcommand{\pp}{~~~.}
\newcommand{\vv}{~~~,}
\newcommand{\nn}{\nonumber}

\begin{document}

\renewcommand{\thefootnote}{\alph{footnote}}
  
\title{
 GROWING NEUTRINO COSMOLOGY}

\author{CHRISTOF WETTERICH}

\address{ Institut  f\"ur Theoretische Physik, Universit\"at Heidelberg,
  Philosophenweg 16\\
 Heidelberg, D-69120, Germany\\
 {\rm E-mail: c.wetterich@thphys.uni-heidelberg.de}}

  \centerline{\footnotesize and}

\author{VALERIA PETTORINO}

\address{Institut  f\"ur Theoretische Physik, Universit\"at Heidelberg, Philosophenweg 16, Heidelberg, D-69120, Germany}

\abstract{In growing neutrino models, the neutrino mass increases in time and stops the dynamical evolution of a dark energy
scalar field, thus explaining the `why now' problem. A new attractive force, mediated by the `cosmon' scalar field, makes non relativistic neutrino form lumps on the scales of
superclusters and beyond. Nonlinear neutrino lumps are predicted to form at redshift $z \approx 1$ and, if observed, could be an indication for a new attractive 
force stronger than gravity.}
\normalsize\baselineskip=15pt
%\section{Growing neutrino cosmologies}
\vspace*{1cm}

It has been recently proposed
\cite{amendola_etal_2007} - \cite{mota_etal_2008} that a growing mass of the neutrinos
may play a key role in stopping the dyna\-mical evolution of the dark energy
scalar field, the cosmon  \cite{wetterich_1988} - \cite{doran_etal_2007}. In these models, the onset of accelerated expansion is triggered
by neutrinos becoming non relativistic. For late cosmology, $z \lesssim 5$, the
overall cosmology is very similar to the usual $\Lambda$CDM concordance model with a cosmological constant, such that the equation of state
of dark energy is close to $w = -1$ and the expansion of the universe
accelerates.
In ``growing neutrino cosmologies'' the present dark energy density $\rho_h(t_0)$ can be expressed in terms of the present average neutrino mass $m_\nu(t_0)$
\be 
\label{rho_h}
\rho_h(t_0)^{1/4} = 1.27 \left(\frac{\gamma m_\nu(t_0)}{eV}\right)^{1/4} 10^{-3} eV
\ee
Here $\gamma$ is a dimensionless ratio of couplings that should be of order one and will be specified later. The coincidence with the observed dark energy density $\rho_h(t_0)^{1/4} = 2 \cdot 10^{-3}$ eV is only possible since the neutrino mass is in the eV (or sub-eV) range. In these models the neutrinos are also singled out by a coupling to the cosmon which is substantially larger than the cosmon coupling to quarks and leptons. 

This particular role of the neutrinos can be clearly linked to the particular way how neutrino-masses are generated in a framework of unified theories.
Indeed, the standard model of electroweak interactions involves only left handed neutrinos such that no renormalizable mass term for the neutrinos is compatible with the gauge symmetry. As a further consequence of the gauge symmetry the difference between baryon number $B$ and lepton number $L$ is conserved by all renormalizable interactions. Neutrino masses can only arise from effective dimension five operators which involve two powers of the vacuum expectation value of the Higgs doublet, $d\approx 175$ GeV. They are suppressed by the inverse power of a large mass scale $M_{B-L}$ which is characteristic for $B-L$ violating effects within possible extensions beyond the standard model. 

The characteristic size of the neutrino masses, $m_{\nu,i}=H_id^2/M_{B-L}$, involves appropriate combinations of dimensionless couplings, $H_i$. Consistency with the observed oscillations requires for the mass of at least one neutrino $m_{\nu,i}\gtrsim 0.05$ eV. For $H_i$ of the order one this implies an upper bound $M_{B-L}\lesssim 6\cdot 10^{14}$  GeV \cite{CWN}. It is notable that this bound is lower than a possible scale of grand unification,  $M_{GUT}\approx 10^{16}$ GeV. This difference in scales is further enhanced if the $H_i$ are smaller than one or if the heaviest neutrino mass is larger than $0.05$ eV. We will assume here a ratio 
$M_{B-L}/M_{GUT}\gtrsim\sqrt{m_s/m_t}\approx 1/30$ as suggested by the necessity of $SU(4)_C$-breaking mass terms for the second generation quarks \cite{CWN}. This requires $H_i$ of the order one or larger, which may not seem very natural if the neutrino masses arise from an induced triplet (see below) and are in the range above $0.1$ eV. In this note we will propose a dynamical mechanism where $H$ is driven to large values in the course of the cosmological evolution.

This mechanism will lead to a fast increase of the neutrino mass from a generic value $m_\nu\lesssim 10^{-3}$ eV to its present substantially larger value. This increase is due to the time evolution of a scalar field - the cosmon - which changes its value even in the present cosmological epoch. The growing neutrino mass has rather dramatic consequences for cosmology. It essentially stops the cosmological evolution of the cosmon and triggers an accelerated expansion of the Universe, thus realizing the ``growing matter'' scenario \cite{amendola_etal_2007}. The increase of the neutrino mass acts as a cosmological clock or trigger for the crossover to a new cosmological epoch. 

The evolution of the cosmon field stops close to a value $\varphi_t$ which is characteristic for the transition between the two different cosmological epochs. This value does not correspond to a minimum of the effective potential $V(\varphi)$ for this scalar field. It is rather selected by a cosmological event, namely the sudden increase of the neutrino masses. The almost constant asymptotic value of the dark energy is given by $V(\varphi_t)$. It is determined by a ``principle of cosmological selection'' rather than by the properties of the vacuum. 

The most general mass matrix for the three light neutrinos reads \cite{CWN}
\be\label{1}
M_\nu=M_DM^{-1}_RM^T_D+M_L.
\ee
The first term accounts for the seesaw mechanism \cite{SS}. It involves the mass matrix for heavy ``right handed'' neutrinos, $M_R$, and the Dirac mass term $M_D=h_\nu d$. The second term accounts for the ``cascade'' or ``induced triplet'' mechanism \cite{MW}
\be\label{2}
M_L=h_L\zeta\frac{d^2}{M^2_t}.
\ee
Here a small expectation value of a heavy $SU(2)_L$-triplet field with mass $M_t$ is induced by a cubic coupling $\zeta$ involving the triplet and two powers of the Higgs-doublet. The triplet carries two units of lepton number such that $\zeta\sim M_{B-L}$. In view of the repetition of the gauge hierarchy $(d/M_{GUT})$ in the respective size of the doublet and triplet expectation values we may call this the ``cascade mechanism''. (The cascade mechanism is often called ``seesaw II'', which seems not the most appropriate name since no diagonalization of a mass matrix with large and small entries is involved, in contrast to the first term in eq. (\ref{1}).) For simplicity we will neglect here the generation structure $(M_\nu$ and $M_L$ are $3\times 3$ matrices) and associate $m_\nu$ with the average neutrino mass 
\be\label{3}
m_\nu=\frac{h^2_\nu d^2}{m_R}+\frac{h_L\zeta d^2}{M_t^2}.
\ee
With $m_R=\sigma M_{B-L}~,~h_L\zeta=\kappa M_{B-L}$ the dimensionless combination $H$, defined by $m_\nu=H d^2/M_{B-L}$, obeys $H=h^2_\nu/\sigma+\kappa M^2_{B-L}/M^2_t$. Discarding large dimensionless couplings $h_\nu,\kappa$ large values of $H$ require small $\sigma$ or small $M^2_t/M^2_{B-L}$. We will realize here the second alternative by a time dependent $M_t$, but a similar mechanism with time dependent $m_R$ is also possible. 

The generic size of the triplet mass is $M_t\approx M_{GUT}$. As a key feature of our scenario we assume that $M_t$ depends on the value of the cosmon field $\varphi$,
\be\label{4}
M^2_t=c_tM^2_{GUT}
\left[1-\frac{1}{\tau}\exp\left(-\epsilon\frac{\varphi}{M}\right)\right],
\ee
with $c_t$ and $\tau$ of the order one, $\tau>1$, and $M$ the reduced Planck mass. For $\epsilon<0$ the triplet mass decreases with increasing $\varphi$ and has a zero, $M_t(\varphi_t)=0$, for 
\be\label{5}
\frac{\varphi_t}{M}=-\frac{\ln\tau}{\epsilon}.
\ee
In consequence, the neutrino mass increases fast when $\varphi$ approaches $\varphi_t$
\be\label{6}
m_\nu(\varphi)=\bar m_\nu 
\left\{ 1-\exp\left[-\frac{\epsilon}{M}(\varphi-\varphi_t)\right]\right\}^{-1}.
\ee
Here we have replaced the parameter $\tau$ by $\varphi_t$ and we neglect the seesaw contribution, which is subleading for the range of $\varphi$ near $\varphi_t$ relevant for our discussion. The parameter $\bar m_\nu$ is given by $\bar m_\nu=(\kappa/c_t)(M_{B-L}/M^2_{GUT})d^2$, with a characteristic size $\bar m_\nu\approx 3\cdot 10^{-5}$ eV for $\kappa/c_t=1/3~,~M_{B-L}/M_{GUT}\approx 1/30$. 

For $\varphi$ near $\varphi_t$ we can approximate
\be\label{7}
m_\nu(\varphi)=\frac{\bar m_\nu M}{\epsilon(\varphi-\varphi_t)}.
\ee
Only this range will be relevant for the quantitative discussion of cosmology below. We observe that the detailed form of the $\varphi$-dependence of $M_t$ is actually not important. It is sufficient that $M^2_t(\varphi)$ crosses zero for $\varphi=\varphi_t$ and admits a Taylor expansion at this point. The neutrino mass depends only on the two effective parameters appearing in eq. (\ref{7}), namely $\varphi_t/M$ and $\bar m_\nu/\epsilon$. Since only a small range of $\varphi$ near $\varphi_t$ plays a role we can neglect the $\varphi$ dependence of all particle physics parameters except for $M_t$ or $m_\nu$. Similar scenarios can be realized by a $\varphi$-dependent mass of the singlet neutrinos $m_R(\varphi)$.

Before discussing cosmology we also have to specify the dynamics of the cosmon field as determined by a Lagrangian $\sim\frac12\partial^\mu\varphi\partial_\mu\varphi+V(\varphi)$. We choose an exponential cosmon potential
\be\label{8}
V(\varphi)=M^4\exp\left(-\alpha\frac{\varphi}{M}\right).
\ee
Cosmology will therefore depend on three parameters, $\alpha,\varphi_t/M$ and $\bar m_\nu/\epsilon$. In addition, the matter density at some initial time, $\rho_M(t_{eq})$, can be mapped into today's value of the Hubble parameter $H_0$. In early cosmology the neutrino mass is negligible and neutrinos behave as a relativistic fluid. Their number density is fixed, as usual, by the physics of decoupling as described by the ratio of effective neutrino and photon temperatures $T_\nu/T_\gamma$. In this early period the cosmological evolution depends only on $\alpha$ and is described by a scaling solution \cite{wetterich_1988} with a constant small fraction of dark energy density
\be\label{9}
\Omega_{h,e}=\frac{n}{\alpha^2},
\ee
with $n=3(4)$ for the matter (radiation) dominated epoch.

However, the neutrino mass grows with increasing 
\be\label{9A}
\varphi=\varphi_0+(2M/\alpha)\ln(t/t_0).
\ee
The scaling period ends once the neutrinos become non-relativistic. Then the cosmon-neutrino coupling influences the field equation for the cosmon \cite{CWQ2}, \cite{fardon_etal_2004}
\ba\label{10}
\ddot\varphi+3H\dot\varphi&=&-\frac{\partial V}{\partial\varphi}+
\frac{\beta(\varphi)}{M}(\rho_\nu-3p_\nu),\\
\beta(\varphi)&=&-M\frac{\partial}{\partial\varphi}\ln m_\nu(\varphi)=\frac{M}{\varphi-\varphi_t}.\nn
\ea
Here $\rho_\nu$ and $p_\nu$ are the neutrino energy density and pressure, obeying
\ba\label{11}
\dot \rho_\nu+3H(\rho_\nu+p_\nu)&=&-\frac{\beta(\varphi)}{M}(\rho_\nu-3p_\nu)\dot\varphi\nn\\
&=&-\frac{\dot\varphi}{\varphi-\varphi_t}(\rho_\nu-3p_\nu).
\ea
The r.h.s of eq. (\ref{11}) accounts for the energy exchange between neutrinos and the cosmon due to the varying neutrino mass \cite{CWQ2}. We observe $\beta(\varphi)<0$ for the range $\varphi<\varphi_t$ where $\varphi$ increases towards $\varphi_t$. The effective coupling $\beta$ diverges for $\varphi\to\varphi_t$ and can therefore become very large for $\varphi$ near $\varphi_t$. This effect stops the evolution of $\varphi$ which approaches the value $\varphi_t$ arbitrarily close but cannot cross it. As a consequence, the potential energy approaches a constant, $V(\varphi)\to V_t=V(\varphi_t)$, which acts similar to a cosmological constant and causes the accelerated expansion. As $\varphi$ approaches $\varphi_t$ the kinetic energy $\dot\varphi^2/2$ must vanish asymptotically. Therefore the equation of state for the cosmon will approach the value $w_\varphi=-1$, and the combined equation of state for the cosmon and neutrinos approaches $w=-V_t/(V_t+2\rho_\nu)$. In this model $\rho_\nu/V_t$ vanishes asymptotically, such that $w\to-1$.

%\subsection{Background}
A crucial ingredient in this model is the dependence of the neutrino mass on
the cosmon field $\varphi$, as encoded in the dimensionless cosmon-neutrino
coupling $ \beta(\varphi) \equiv - \frac{d \ln{m_\nu}}{d \varphi}.$ For increasing $\varphi$ and $\beta < 0$ the neutrino mass increases with
time $ m_{\nu} = \bar{m}_{\nu} e^{-{\tilde{\beta}}(\varphi) \varphi},$ where
$\bar{m}_{\nu}$ is a constant and $\beta = \tilde{\beta} + \partial
\tilde{\beta}/\partial \ln{\varphi}$. Different particle physics models for the growing neutrino scenario will correspond to different functions $\beta(\varphi)$. We will concentrate in the following on the simplest case \cite{amendola_etal_2007} of constant $\beta$ where the parameter $\gamma$ in Eq.~(\ref{rho_h}) reads
\be \label{c1}
\gamma = -\frac{\beta}{\alpha} \vv
\ee
where we recall that $\beta \sim 1$ corresponds to a cosmon mediated interaction for neutrinos with gravitational strength. 
For a comparison, the particular model described above has 
\be \label{c2} \gamma = -\frac{\epsilon}{\alpha} \frac{m_\nu(t_0)}{\bar{m}_\nu} \pp \ee
We have checked that a varying $\beta(\varphi)$ leads to qualitatively similar cosmologies as for constant $\beta$.
From now on we normalize the cosmon field $\varphi$ in units of the reduced Planck mass $M = (8 \pi G_N)^{-1/2}$.

The homogeneous energy density and pressure of the scalar field $\varphi$ are defined
in the usual way \cite{mota_etal_2008}. 
It will be useful to express the conservation equations for dark energy and neutrino densities as \cite{wetterich_1995} \cite{amendola_2000} 
\bea \label{cons_phi} \rho_{\varphi}' &=& -3 {\cal H} (1 + w_\varphi) \rho_{\varphi} +
\beta(\varphi) \varphi' (1-3 w_{\nu}) \rho_{\nu} \vv \\
\label{cons_gr} \rho_{\nu}' &=& -3 {\cal H} (1 + w_{\nu}) \rho_{\nu} - \beta(\varphi) \varphi' (1-3 w_{\nu}) \rho_{\nu}
\vv \nonumber \eea 
where prime indicates derivative with respect to conformal time, ${\cal H} \equiv a'/a$ is the Hubble parameter, $w_\varphi \equiv p_\varphi/\rho_\varphi$ and $w_\nu \equiv p_\nu/\rho_\nu$ are the equations of state of the cosmon field and neutrinos respectively.
The sum of the energy momentum tensors for neutrinos and the cosmon is conserved,
 but not the separate parts. We neglect a possible cosmon coupling
to Cold Dark Matter (CDM), so that $\label{cons_cdm} \rho_c' = -3 {\cal H} \rho_c $.

\begin{figure}[ht]
%\begin{picture}(185,200)(0,0)
\centering
\includegraphics[width=75mm,angle=0.]{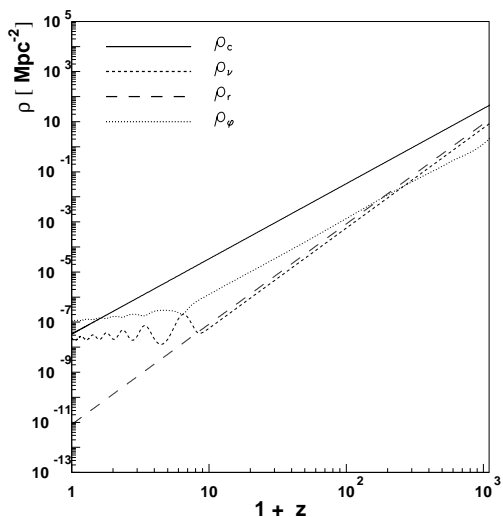}% Here is how to import EPS art
%\end{picture}
\caption{Energy densities of neutrinos (dashed), cold dark matter (solid), dark energy (dotted) and photons (long dashed) are plotted vs redshift. For all plots we take a constant $\beta = -52$, with $\alpha = 10$ and large neutrino mass $m_\nu = 2.11$ eV.}
\label{fig_1}
\vspace{0.5cm}
\end{figure}

For a given potential (\ref{8}) the evolution equations for the different species can be numerically integrated, giving the background
evolution shown in Fig:~\ref{fig_1} (for constant $\beta$) \cite{amendola_etal_2007}. The initial
pattern is a typical early dark energy model, since neutrinos are still relativistic and almost
massless. Radiation dominates until matter radiation equality, when CDM takes over. Dark
energy is still subdominant and falls into the attractor provided by the
exponential potential (see \cite{wetterich_1995} \cite{amendola_2000} for
details). As the mass of the neutrinos increases with time, the term
$\sim \beta \rho_\nu$ in the evolution equation for the cosmon (\ref{cons_phi})
starts to play a more significant role, kicking
$\varphi$ out of the attractor as soon as neutrinos become non-relativistic. This resembles the effect of the coupled dark matter component in \cite{huey_wandelt_2006}. Subsequently, small decaying oscillations characterize the $\varphi - \nu$ coupled fluid
and the two components reach almost constant values. The va\-lues of the energy densities
today are in agreement with observations, once the precise crossing time for the end of the scaling 
solution has been fixed by an appropriate choice of the coupling $\beta$. At
present the neutrinos are still subdominant with respect to CDM, though in the future
they will take the lead (see
\cite{amendola_etal_2007} for details on the future attractor solution for constant $\beta$).
 
\begin{figure}[ht] 
\centering
\begin{minipage}{70mm}
\begin{center}
%\begin{picture}(185,200)(20,0)
\includegraphics[width=78mm,angle=0.]{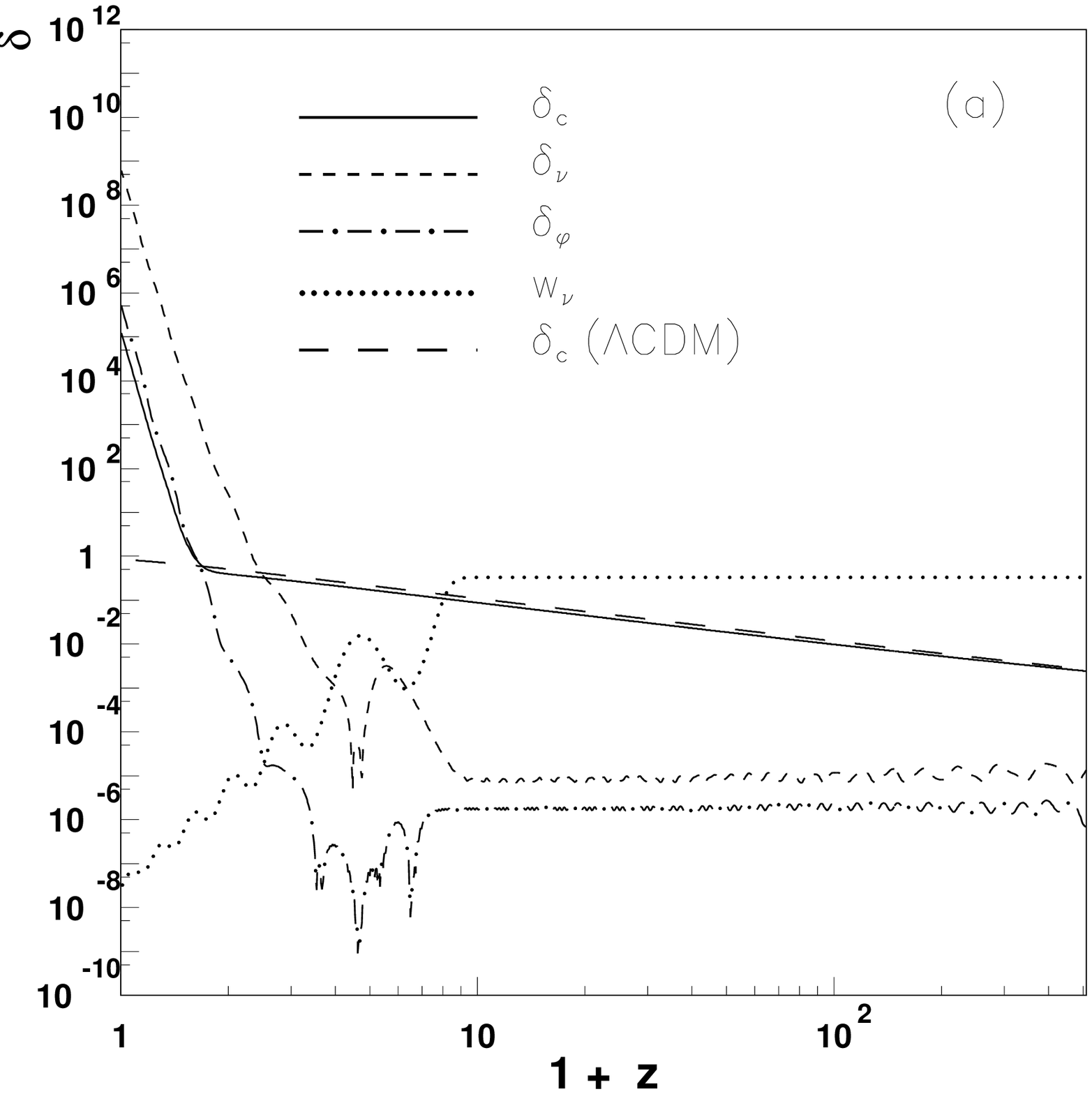}% Here is how to import EPS art
%\end{picture}
\label{fig_2a}
\end{center}
\end{minipage}
\hspace{4mm}
\begin{minipage}{70mm}
\begin{center}
%\begin{picture}(185,200)(20,0)
\includegraphics[width=78mm,angle=0.]{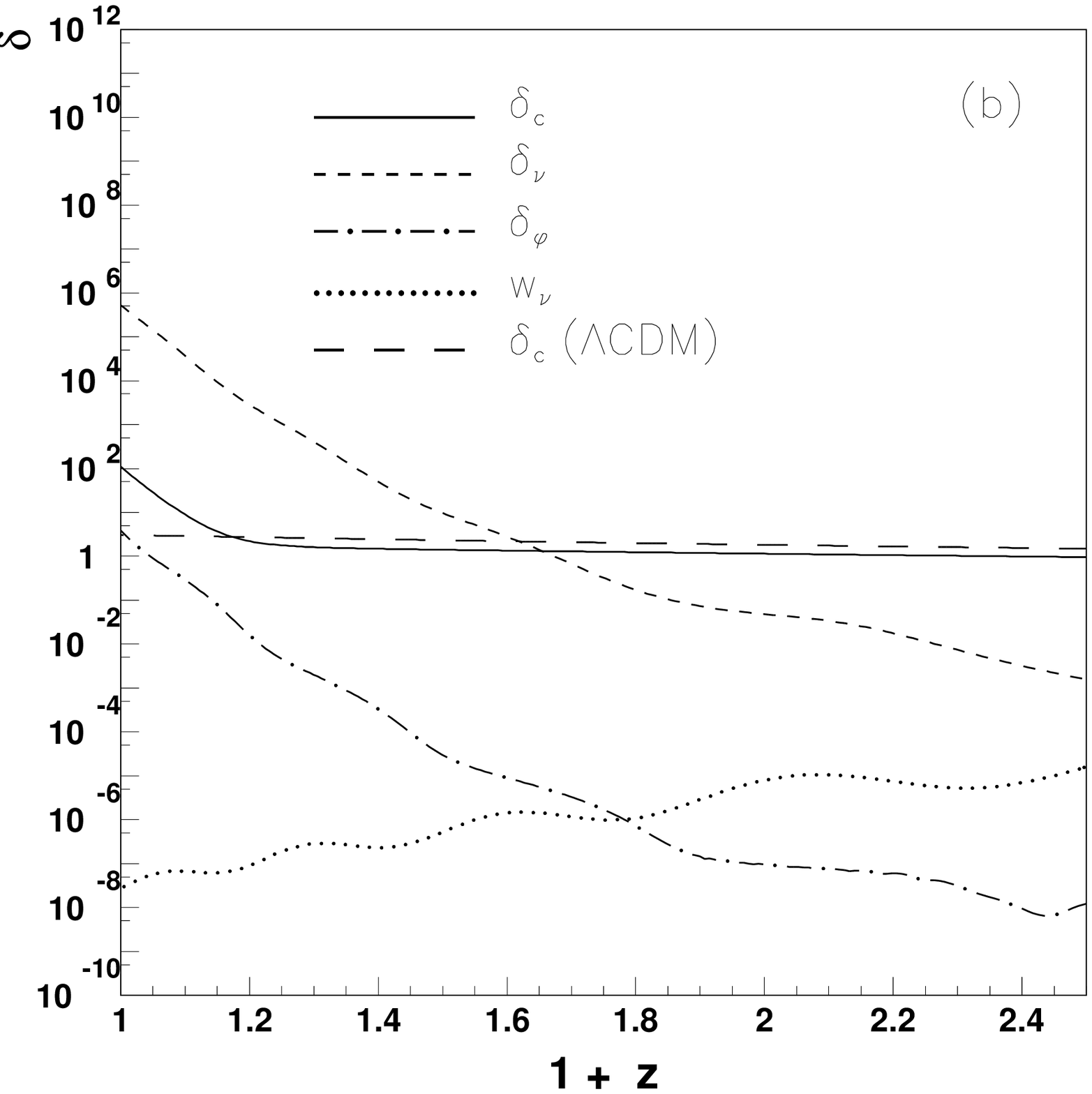}% Here is how to import EPS art
%\end{picture}
\label{fig_2b}
\end{center}
\end{minipage}
\vspace{4mm}
\caption{Longitudinal density perturbation for CDM (solid), $\nu$ (dashed) and $\varphi$ (dot-dashed) vs redshift for $k = 0.1 h/Mpc$ (upper panel) and $k = 1.1 h/Mpc$ (lower panel, $\lambda = 8 Mpc$). The neutrino equation of state (dotted) is also shown. The long dashed line is the reference $\Lambda$CDM.}\label{fig_2}
\end{figure}

An efficient stopping of the cosmon evolution by the relatively small energy
density of neutrinos needs a cosmon-neutrino coupling that is somewhat larger
than gravitational strength. This is similar to mass varying neutrino
models \cite{fardon_etal_2004} \cite{afshordi_etal_2005}, even though the coupling is much smaller and effective at late times, with a corresponding light cosmon field.
The enhanced attraction between
neutrinos leads to an enhanced growth of neutrino fluctuations, once the
neutrinos have become non-relativistic \cite{afshordi_etal_2005} \cite{bjaelde_etal_2007}. In view of
the small present neutrino mass, $m_\nu(t_0) < 2.3$ eV, and the time
dependence of $m_\nu$, which makes the mass even smaller in the past, the time
when neutrinos become non-relativistic is typically in the recent history of
the universe, say $z_R \approx 5$. Neutrinos have been free streaming for $z >
z_R$, with a correspondingly large free streaming length. Fluctuations on
length scales larger than the free streaming length are still present at
$z_R$, and they start growing for $z < z_R$ with a large growth rate. This
opens the possibility that neutrinos form nonlinear lumps
\cite{amendola_etal_2007} \cite{brouzakis_etal_2007} on supercluster scales, thus
opening a window for observable effects of the growing neutrino scenario.

Neutrino perturbations indeed grow non-linear in
these models. Non-linear neutrino structures form at redshift $z \approx 1$ on
the scale of superclusters and beyond. One may assume that these structures
later turn into bound neutrino lumps of the type discussed in
\cite{brouzakis_etal_2007}. 
Our investigation is here limited to linear perturbations. We can therefore provide a reliable estimate for the time when the first fluctuations become non-linear. For later times, it should only be used to give qualitative limits.

%\subsection{Linear perturbations}
The evolution equations for linear perturbations (in Fourier space), in Newtonian gauge (in which the non diagonal metric perturbations are fixed to zero) \cite{kodama_sasaki_1984}, is fully described in \cite{mota_etal_2008}. We numerically compute the linear density
perturbations both using a modified version of CMBEASY \cite{cmbeasy} and, independently, a modified version of CAMB \cite{lewis_etal_2000}. We plot the density fluctuations $\delta_i$ as a function of
redshift for a fixed $k$ in Fig:~\ref{fig_2}.
The neutrino equation of state is also shown, starting from
$1/3$ when neutrinos are relativistic and then decreasing to its present value
when neutrinos become non relativistic. The turning point marks the time at
which neutrino perturbations start to increase. At the scale of $k = 0.1 h/Mpc$
(corresponding to superclusters scales) shown in Fig:~\ref{fig_2}$a$, neutrino
perturbations eventually overtake CDM perturbations and even force $\varphi$
perturbations to increase as well, in analogy with dark 
energy clustering expected in \cite{perrotta_baccigalupi_2002} \cite{pettorino_baccigalupi_2008} within scalar tensor 
theories. Notice, however, that the scale at which
neutrinos form nonlinear clumps depends on the model parameters, in particular
the coupling $\beta$, the potential parameter $\alpha$ and
the present neutrino mass. Those are related to the neutrino
free-streaming length, the range of the cosmon
field and its mass. A detailed investigation of the parameter space will be performed in future work, but we mention that the model \cite{wetterich_2007} with varying $\beta$ gives qualitatively similar results.

\begin{figure}[ht]
\centering
\includegraphics[width=75mm,angle=0.]{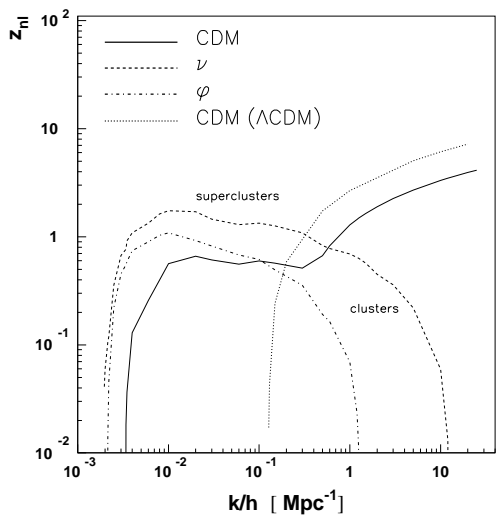}% Here is how to import EPS art
\caption{Redshift of first non linearities vs the wavenumber $k$ for CDM (solid), $\nu$ (dashed) and $\varphi$ (dot-dashed). We also plot CDM for a reference $\Lambda$CDM model (dotted).\label{fig_4}}
\end{figure}

We emphasize again that once the neutrinos form strong nonlinearities - neutrino lumps - one expects that nonlinear effects substantially slow down the increase of $\delta_\nu$ and even stop it. The magnitude of the CDM-dragging by neutrinos is therefore not shown - it might be much smaller than visible in Fig:~\ref{fig_2}. These remarks concern the quantitative interpretation of all the following figures, which are always computed in the linear approximation.
Nevertheless, the linear approximation demonstrates well the mechanisms at work. 

In Fig:~\ref{fig_4} we plot the redshift $z_{nl}$ at which CDM,
neutrinos and $\varphi$  become nonlinear as a function of the
wavenumber $k$. The case of CDM in the concordance $\Lambda$CDM model with the same present value of $\Omega_{\varphi}$ and for massless neutrinos is also
shown for reference (dotted line). The
redshift $z_{nl}$ roughly measures when nonlinearities first appear by evaluating the
time at which $\delta(z_{nl}) = 1$ 
for each species. The curves in Fig:~\ref{fig_4} are obtained in the linear approximation, such that only the highest curves are quantitatively reliable.  This concerns CDM for large $k$ and neutrinos for small $k$. The subleading components are influenced by dragging effects and may be, in reality, substantially lower. 

We can identify four regimes: i) At very big scales (larger than superclusters) the universe is
homogeneous and perturbations are still linear today.
ii) The range of length scales going from 
%0.6 h/Mpc   ;  0.002 h/Mpc h = 0.72 
$14.5$ Mpc to about $4.4 \times 10^3$ Mpc appears to be
highly affected by the neutrino coupling in growing matter scenarios: neutrino
perturbations are the first ones to go nonlinear and neutrinos seem to form
clumps in which then both the scalar field and CDM could fall into. Note that the
effect of the neutrino fluctuations on the gravitational potential induces CDM to
cluster earlier with respect to
the concordance $\Lambda$CDM  model, where CDM is still linear at scales above $\sim$87 Mpc.
iii) For lengths included in the range between
% 0.6 h/Mpc and 10 h/Mpc
$0.9$ Mpc and $14.5$ Mpc, CDM takes
over. That is in fact expected since neutrinos start to approach the free streaming scale. In this regime CDM drags neutrinos, and this effect may be overestimated in the linear approximation.
Notice that in our model CDM clusters later than it would do in $\Lambda$CDM. 
There are two reasons for this effect. At early times, the presence of a homogeneous component of early dark energy, 
$\Omega_{\varphi} \sim 3/\alpha^2$, implies that $\Omega_m$ is somewhat smaller than one and therefore clustering is slower \cite{doran_etal_2001}.
At later times, $\Omega_m$ is smaller than in the $\Lambda$CDM model since for the same $\Omega_{\varphi}$, part of $1-\Omega_{\varphi} = \Omega_m + \Omega_\nu$
is now attributed to neutrinos. In consequence massive neutrinos reduce structure at smaller scales when they do not contribute to the clumping. The second effect is reduced for a smaller present day neutrino mass. 
iv) Finally, at very small scales (below clusters), CDM becomes highly non linear and
neutrinos enter the free streaming regime, their perturbations do not growth and remain
inside the linear regime.

Maximum neutrino clustering occurs on supercluster scales and one may ask about observable consequences. First of all, the neutrino clusters could have an imprint on the CMB-fluctuations. Taking the linear approximation at face value, the ISW-effect of the particular model presented here would be huge and strongly ruled out by observations. However, non-linear effects will substantially reduce the neutrino-generated gravitational potential and the ISW-effect. Further reduction is expected for smaller values of $\beta$ (accompanied by smaller $\alpha$). It is well conceivable that realistic models lead to an ISW-effect in a range interesting for observations.
A second possibility concerns the detection of nonlinear structures at very large length scales. Such structures can be found via their gravitational potential, independently of the question if neutrinos or CDM source the gravitational field. Very large nonlinear structures are extremely unlikely in the $\Lambda$CDM concordance model. An establishment of a population of such structures, and their possible direct correlation with the CMB-map \cite{rudnick_etal_2007} - \cite{giannantonio_etal_2008}, could therefore give a clear hint for ``cosmological actors'' beyond the $\Lambda$CDM model. For any flat primordial spectrum the gravitational force will be insufficient to produce large scale clumping, which could thus be an indication for a new attractive force stronger than gravity - in our model mediated by the cosmon.

Let us end with the remark that there is a chance that the time variation of the neutrino mass could even be detected. Indeed, our understanding of structure formation and other cosmological features place strong upper bounds on the neutrino masses in early cosmology, say $z \gtrsim 5$. One may argue about the precise location of this bound, but a neutrino mass of $0.5$ eV would certainly have left a strong imprint on cosmology which has not been observed. In consequence, if the direct searches for a neutrino mass or the neutrinoless double beta decay indicate a neutrino mass larger than $0.5$ eV, this could be interpreted as a strong signal in favor of a growing neutrino mass.  

%\section{Acknowledgements}

\end{document}